# Moiré Topology in Twisted Structures with Noncollinear Spin-Orbit Coupling

*Xilong Xu,*[1] *Liangtao Peng,*[1] *Shaffique Adam*,*[1,2] *Li Yang*,*[1, 2]

[1]Department of Physics, Washington University in St. Louis, St. Louis, Missouri 63130, USA

[2]Institute of Materials Science and Engineering, Washington University in St. Louis, St. Louis, Missouri 63130, USA

* Email: lyang@physics.wustl.edu, shaffique@wustl.edu

**ABSTRACT**

Moiré superlattices provide a powerful platform for flat bands and correlated topological phases, yet most established examples rely on valley-contrasting Berry curvature in hexagonal lattices. Here, we propose a different route to achieve topological moiré minibands based on noncollinear spin–orbit coupling (SOC) in centrosymmetric type-II SOC bilayers. Interlayer hybridization opens local pseudogaps and produces sharply localized Berry curvature, which twisting reconstructs into isolated topological minibands without requiring valley degrees of freedom or hexagonal symmetry. We demonstrate this mechanism in tetragonal Dresselhaus-SOC $HgI_2$, where a Lieb-like moiré potential yields topological flat bands. To improve miniband isolation, we develop a physics-informed machine-learning surrogate that identifies stronger SOC as a key design principle and guides the replacement of Hg by Pb. The resulting $PbI_2$ minibands are narrower and better isolated, supporting correlation-driven magnetism and tunable quantum spin Hall and Chern insulating phases, thereby providing an optimized material realization for experimental exploration.

Moiré superlattices formed by twisting or lattice-mismatching two-dimensional (2D) van der Waals (vdW) materials have emerged as a powerful and highly tunable platform for engineering exotic quantum states.[1–7] In the small-angle regime, moiré modulation can produce exceptionally flat electronic bands,[8–10] providing an ideal setting in which band topology and electron–electron interactions are simultaneously amplified.[11–18] This interplay has generated a rapidly growing family of emergent quantum phases.[19–23] An important microscopic origin of moiré topology lies in the quantum geometry of the parent bands, especially the momentum-space distribution of Berry curvature. In established platforms, nontrivial moiré minibands are often inherited from intrinsic Berry-curvature hotspots already present in the constituent monolayers.[15–18,24–28] For example, in transition-metal dichalcogenides (TMDs), crystal symmetry and intrinsic SOC generate valley-contrasting Berry curvature sharply concentrated near the band edges, and moiré reconstruction magnifies these geometric features into topological minibands.[29–35] This mechanism has underpinned many influential discoveries, but it also leaves open a more fundamental question: is valley physics truly essential for moiré topology, or can topological minibands arise more generally from SOC-driven quantum geometry in materials without valley degrees of freedom or preexisting valley Berry curvature?

Noncollinear spin–orbit coupling, including Dresselhaus and Rashba types, offers a distinct and largely unexplored starting point.[36–40] In Rashba and Dresselhaus systems, spin-momentum locking can generate avoided crossings with strong Berry-curvature responses once appropriate hybridization gaps are opened.[41–43] Our previous work on twisted type-II Rashba homobilayers demonstrated one realization of this idea by showing that hexagonal BiTeI-based R-2 systems can host tunable topological moiré flat.[44] However, that study focused on a specific Rashba spin helicity and a threefold moiré geometry. It therefore remained unclear whether the underlying mechanism is unique to type-II Rashba systems or can serve as a more general route across different noncollinear SOC textures, lattice symmetries, and material classes. This question is particularly important for Dresselhaus SOC, which belongs to a symmetry setting distinct from Rashba SOC. In tetragonal and square-lattice materials, Dresselhaus SOC can generate spin textures and moiré potential landscapes that cannot be simply reduced to the Rashba case. Therefore, extending the type-II SOC moiré concept from Rashba to Dresselhaus systems and further including many-electron correlations provides a critical test of the generality of noncollinear-SOC-driven moiré topology.[45,46]

Here, we show that noncollinear SOC provides a general seed for topological moiré minibands, with type-II Rashba and type-II Dresselhaus bilayers representing two symmetry-distinct realizations, as illustrated in **Fig. 1**. Type-II Rashba (R-2) or Dresselhaus SOC (D-2) refers to a centrosymmetric bilayer constructed from two noncentrosymmetric monolayers with opposite polar axes or symmetry-related orientations. The central mechanism is that interlayer hybridization in centrosymmetric type-II SOC bilayers opens local pseudogaps and produces sharply localized

Berry curvature near avoided crossings. Twisting then introduces a long-wavelength moiré potential that slices, redistributes, and reconstructs these Berry-curvature hotspots into isolated topological minibands. Importantly, this mechanism does not require valley degeneracy, hexagonal symmetry, or topological monolayer bands.

We demonstrate this principle in twisted tetragonal Dresselhaus-SOC $HgI_2$, where the square-lattice symmetry produces a Lieb-like moiré potential landscape, providing a non-hexagonal realization of topological moiré minibands. To move beyond a single-material example, we further develop a physics-informed ML surrogate model to identify the microscopic parameters that favor isolated topological flat bands and use the resulting SOC-based design rule to guide Pb-based material optimization. Finally, Hartree–Fock calculations show that these optimized minibands are not merely topological single-particle bands, but can also support interaction-driven magnetic and topological insulating phases. Together, these results establish noncollinear-SOC moiré systems as a general platform for engineering correlated topology beyond the current valley-based paradigm.

*Dresselhaus-SOC structure*: **Figure 1** schematically summarizes the mechanism, from the construction of type-II noncollinear-SOC bilayers to the formation of Berry-curvature hotspots and their reconstruction into topological moiré minibands upon twisting. The confirmation of topological flat bands of twisted type-II Rashba homobilayers is presented in supporting information (Section I in SI). In the main text, we therefore focus on demonstrating this design principle in 2D Dresselhaus systems. Dresselhaus SOC is a fundamental spin-orbit interaction arising from bulk inversion asymmetry inherent to the lattice.[47] The general Hamiltonian around the band edge can be described by:[48,49]

$$H = \frac{\hbar^2 k^2}{2m^*} + \alpha_D\left(k_x\sigma_x - k_y\sigma_y\right) \quad (1)$$

Here, $\sigma$ represents the spin Pauli metrices. The parameter $\alpha_D$ represents the strength of the Dresselhaus SOC, which depends on the atomic SOC as well as the degree of spatial inversion symmetry breaking. The Dresselhaus SOC effect exhibits a gapless Dirac cone band structure with the characteristic helical spin polarizations, as schematically illustrated in **Figs. 2 (a1)** and **(a2)**. This effect has been widely observed in non-centrosymmetric materials and plays a key role in spin splitting and spin momentum locking of electronic bands. These features make Dresselhaus SOC highly relevant for spintronic applications, such as spin-field-effect transistors, spin–orbit torque switching, and electrically controlled spin transport, as well as for spin-dependent phenomena including the spin Hall effect and topological states.[47,50,51]

However, in conventional gapless Rashba or Dresselhaus systems, the relevant spin-split bands are not isolated, and symmetry constraints often suppress Berry curvature away from band degeneracies. Consequently, these materials have not been directly connected to isolated topological flat bands. Although previous studies have shown that localized Berry curvature can be generated by applying a weak out-of-plane magnetic field to open a pseudo-gap at the original band-crossing points,[41] the use of an external magnetic field is an extrinsic approach and is difficult to scale in a practical manner.

Recently, interlayer engineering has provided a promising route to generate Berry curvature in Rashba- or Dresselhaus-SOC systems without relying on an external magnetic field. This interlayer-engineered SOC configuration is referred to as type-II Rashba or type-II Dresselhaus SOC, abbreviated as R-2 or D-2 SOC, respectively.[52–56] The notation "R-2/D-2" denotes a centrosymmetric bilayer composed of two Rashba- or Dresselhaus-active monolayers whose local SOC fields are opposite between the two layers, as illustrated in **Fig. 1**. In contrast to conventional noncentrosymmetric Rashba/Dresselhaus systems, an R-2/D-2 bilayer restores global inversion symmetry while retaining hidden layer-resolved SOC textures. Taking the D-2 homobilayer as an example, the effective Hamiltonian can be written as:

$$H = \frac{\hbar^2 k^2}{2m^*}\sigma_{L_0} + \alpha_R\left(k_x\sigma_y - k_y\sigma_x\right)\sigma_{L_z} + \alpha_I\sigma_{L_x}\,, \qquad (2)$$

where $\sigma_{L_0}$, $\sigma_{L_z}$, $\sigma_{L_x}$ are defined as the layer Pauli metrices. $\alpha_I$ is the strength of interlayer coupling, and the term, $\alpha_I\sigma_{L_x}$, accounts for the interlayer coupling energy. Importantly, as shown in **Figs. 2(b)** and **2(c)**, a pseudo-gap is created at the crossing point with highly localized berry curvature. In fact, the interlayer interaction essentially works as a pseudo antiferromagnetic field after a unitary transformation of the Hamiltonian Eq. (2). As a result, the intrinsic in-plane Dresselhaus spin texture is biased to the out-of-plane direction around the gap opening area in the D-2 homobilayer, which is shown in **Fig. 2(b2)**.

Building on this mechanism, if these D-2 systems with strongly localized Berry curvature are further subjected to a large-scale periodic modulation, such as a moiré potential, the resulting band folding and gap openings around mini-zone boundaries, could give rise to isolated minibands and subsequent correlated topological phases. In the following, we demonstrate this idea in real material systems. Particularly, while twisted quantum materials have been studied predominantly in hexagonal lattices, other lattice symmetries remain much less explored. More specifically, tetragonal systems are largely absent from the current moiré topology landscape, and topological materials with fourfold symmetry are still rare, motivating our choice of a 2D tetragonal lattice as the starting platform.

**$HgI_2$** is a layered tetragonal material that exhibits a clear Dresselhaus SOC band structure, making it an ideal platform for our study. It has been synthesized as a vdW material with low cleavage energy.[57] In addition, this material has been reported to host pronounced anisotropy and fractional quantum ferroelectricity.[57,58] **Fig. 3(a)** plots the crystal structure of monolayer $HgI_2$. It consists of a central Hg atomic layer sandwiched between two layers of I atoms. Each Hg atom is coordinated by four surrounding I atoms, forming a tetrahedral local environment. Owing to its $S_4$ symmetry, $HgI_2$ does not exhibit net electric polarization. However, the lattice lacks inversion symmetry within the plane, giving rise to a Dresselhaus-type SOC. The band structure of monolayer $HgI_2$ is presented in **Fig. 3(b)**, where the band edges exhibit pronounced spin splitting and symmetry-protected band crossings. The spin texture of VBM in reciprocal space is shown in **Fig. S2**, and it confirms the classical Dresselhaus SOC spin texture.

The D-2 homobilayer can be formed by stacking two monolayer $HgI_2$ with antiparallel direction, as shown in the inset of **Fig. 3(c)**, which restores global inversion symmetry. The band structure of bilayer $HgI_2$ is presented in **Fig. 3(c)**. As expected from the Hamiltonian (Eq. (2)), the Dirac cone is now gapped due to interlayer hopping, forming a pair of parabolic-like energy dispersions. Each band is spin degenerate due to the *PT* symmetry.

**Twisted $HgI_2$ homobilayer:** Next we twist the D-2 $HgI_2$ homobilayer at a commensurate small angle to form moiré superlattices. Here we focus on the valence band edge, and the corresponding effective Hamiltonian is

$$H_{Moiré}(\boldsymbol{r},\boldsymbol{k}) = \begin{pmatrix} \frac{\hbar^2 k^2}{2m^*} + \alpha_D\left(k_x\sigma_x - k_y\sigma_y\right) & \alpha_I(\boldsymbol{r})\sigma_0 \\ \alpha_I(\boldsymbol{r})\sigma_0 & \frac{\hbar^2 k^2}{2m^*} - \alpha_D\left(k_x\sigma_x - k_y\sigma_y\right) \end{pmatrix} + \Delta(\boldsymbol{r}), \tag{3}$$

where $\alpha_I(\boldsymbol{r})$ represents the interlayer interaction with spatial modulation, and $\Delta(\boldsymbol{r})$ denotes the moiré potential. The spatially varying terms $\Delta(r)$ and $\alpha_I(r)$ are constructed from stacking-dependent untwisted bilayer DFT calculations, following the standard continuum-model strategy used in moiré systems. Specifically, we calculate four high-symmetry local stackings, AA, AB, AC, and AD, in the square-lattice D-2 $HgI_2$ bilayer. For each stacking, $\Delta$ is extracted from the shift of the relevant band edge, while $\alpha_I$ is obtained from the band splitting between the first and second *PT*-degenerate band pairs near the band edge. These discrete stacking-dependent values are then fitted to the leading symmetry-allowed moiré reciprocal-lattice harmonics to obtain the continuous functions $\Delta(r)$ and $\alpha_I(r)$. The detailed fitting procedure, Fourier expansion, and the resulting real-space maps of both $\Delta(r)$ and $\alpha_I(r)$ are provided in the Supporting Information.[29,30] **Figure 3(d)** shows the scalar moiré potential $\Delta(r)$ of twisted D-2 $HgI_2$ at $\theta = 3.5°$. Notably, unlike hexagonal lattices with threefold symmetry, the square lattice $HgI_2$ has four specific local stackings and gives rise to a Lieb-like moiré potential landscape. The corresponding spatial modulation of the interlayer tunneling $\alpha_I(r)$, which controls the hybridization gap and Berry-curvature generation, is shown in Fig. S3. Finally, with these parameters determined, solving the Hamiltonian yields the moiré band structure and associated topological properties.

A calculated electronic band structure of twisted D-2 $HgI_2$ homobilayer is presented in **Fig. 3(e)** at a twist angle of 3.5°. As expected, the parent Dresselhaus-split bands are folded into mini BZ and reconstructed into a set of narrow minibands. To determine the topological character of these minibands, we evaluate the $Z_2$ invariant using the Wilson-loop formalism, by tracking the evolution of the hybrid Wannier charge centers (WCCs) for the occupied states. We particularly concentrate on the two pairs of the highest-energy valence minibands because they are well within the practical doping (filling) range. These bands remain separated from all other higher-energy bands by a finite direct gap across the entire mini BZ. A topologically nontrivial phase is signaled by an odd number of crossings between the WCC flow and an arbitrary reference line (corresponding to $Z_2 = 1$). The calculated WCC spectrum of the two pairs of the highest-energy valence minibands are presented in the inset, which shows a clear odd winding of the WCCs, providing unambiguous evidence for the nontrivial $Z_2$ topology and the prediction of the quantum spin Hall effect (QSH). Finally, we have calculated the evolution of the bandwidth and topological phase of second pair valence bands with the twist angle in **Fig. 3(f)**. The bandwidth is around 20 meV within this small angle range 3° - 4° and the critical twist angle for the topological phase transition is about 3.4°.

Despite these promising features, two major challenges remain. First, the miniband width is still relatively large. Compared with twisted TMDs and R-2 homobilayers (e.g., BiTeI), whose miniband widths are approximately 10 meV and 7 meV at these similar twist angles,[44,59] the broader bandwidth in twisted $HgI_2$ weakens many-electron interactions and may suppress the emergence of correlated topological phases. Second, although the first and second minibands are separated from neighboring bands by direct gaps, their energy windows overlap with other minibands, as shown in **Fig. 3(e)**, which prevents the formation of a global spectral gap required for observing quantized transport states. These limitations highlight the need to identify new Dresselhaus materials capable of producing narrower and more isolated moiré minibands.

However, discovering such material candidates is highly nontrivial. The vast number of possible materials, together with the large parameter space of moiré structures, such as interlayer interaction, SOC intensity, and twist angle, makes it difficult to determine how individual parameters control the resulting electronic properties. To address this challenge, we employ a ML approach to systematically explore the high-dimensional parameter space and identify the primary determinants governing moiré band topology.

*ML-assisted physical interpretation*: As summarized in **Fig. 4(a)**, we develop a physics-informed ML model that provides a data-efficient, interpretable, and systematically improvable surrogate for first-principles topological classification, using training data generated from Eq. (3). We consider a twisted bilayer–type system whose low-energy electronic topology is governed by a high-dimensional parameter space. Specifically, the model is parameterized by a 10-dimensional (hexagonal moiré) or 12-dimensional (square moiré) vector:

$$\mathbf{X} = (T_{m,i}, T_{e,i}, m^*, a_0, \theta, \lambda_{SOC}), \tag{4}$$

where $\{T_{m,i}\}$ and $\{T_{e,i}\}$ encode the moiré modulation of effective pseudomagnetic field and interlayer potential terms, $m^*$ is the effective mass, $a_0$ is the lattice constant, $\theta$ is the twist angle, and $\lambda_{SOC}$ is the Rashba or Dresselhaus SOC strength. Except for twist angle, the rest can be obtained for a specific material using *ab initio* methods.

We focus on the two pairs of bands close to band extrema. Topological characterization is performed by computing Wilson loops and the $\mathbb{Z}_2$ invariant. In parallel, we evaluate the minimum direct gaps across the mini–Brillouin zone, such as $\min_{\mathbf{k}}(E_3 - E_2)$ and $\min_{\mathbf{k}}(E_5 - E_4)$, corresponding to two physically relevant band fillings. To ensure that the predicted topological phases are experimentally meaningful, we introduce a 1 meV gap threshold to define a binary "gapped versus gapless" label in addition to the topological invariant itself. This criterion filters out formally topological but practically gapless cases, which may otherwise be classified as positive candidates in high-throughput searches despite being difficult to observe experimentally. This procedure yields a supervised dataset $\{(\mathbf{x}_i, \mathbf{y}_i)\}$ comprising (i) discrete targets, the $Z_2$ indices and gap-existence flags at selected fillings, and (ii) continuous targets, the direct gap magnitudes. For the surrogate model, we employ gradient-boosted decision trees (GBDT) model (see details in Section III of SI), which are well suited for learning nonlinear relationships in high-dimensional parameter spaces arising from continuum moiré Hamiltonians.[60,61]

To obtain a more reliable evaluation of the ML classification performance, we report both the overall accuracy (Acc) and balanced accuracy (BalAcc) (see **Fig. 4(b)** and Section III.B of the SI). Acc measures the fraction of correctly predicted samples among all samples, whereas BalAcc averages the true-positive and true-negative rates and is therefore more informative when the dataset is class-imbalanced.[62,63] The model is trained on a dataset containing more than 8,000 samples, and the evaluations of the ML process are summarized in **Table 1**. The ML model achieves a prediction accuracy above 96% for most classification tasks, while the accuracy for the $Z_2$ index of the second miniband pair decreases to approximately 75%. We attribute this reduced performance to the fact that higher minibands are generally less isolated and more susceptible to band mixing with adjacent bands, making their topological labels more sensitive to small parameter variations. These results demonstrate that the proposed ML framework accurately captures both spectral and topological properties while providing an efficient surrogate for parameter-space exploration.

In a brute-force workflow, each candidate parameter set requires explicit band-structure, gap, and Wilson-loop calculations, making large-scale searches computationally expensive. By contrast, the trained ML model enables rapid prescreening of candidate regions, so that full topological calculations are needed only for a much smaller set of promising structures.

We have analyzed how the trained model exploits the physics-informed features. Along all the features, the SOC strength of $\lambda_{SOC}$ consistently emerges as the most influential feature, with a dominant importance weight and small variance across folds, as shown in **Fig. 4(c)**. Feature importance extracted from the GBDT ensemble and averaged over cross-validation folds is summarized in Table S2 (see details in Section III(c) of SI). This is fully consistent with the central role of Rashba or Dresselhaus SOC in generating gap opening and topological band inversion.

Further insight is provided by the partial dependence plots (PDPs), shown in **Fig. 4(d)**. The predicted probability of a nontrivial phase of first pair bands ($Z_2^2$) increases monotonically with increasing $\lambda_{SOC}$, reflecting SOC-driven enhancement of Berry curvature and effective band inversion at low fillings. In contrast, the response of nontrivial phase of the second pair bands ($Z_2^4$) is much weaker and nearly flat over the same parameter range, as shown in **Fig. 4(d)**. This behavior reflects the increased fragility of topology at higher fillings, where multiple bands participate and topological character depends on a more delicate balance among parameters.

*Isolated twisted topological bands and correlated effect*: To verify the guidance from the ML model, we manually increased the SOC strength to 2.5 times its original value while keeping all other parameters in the effective Hamiltonian of the twisted $HgI_2$ homobilayer (Eq. (3)) unchanged. As shown in **Fig. 5(a)**, in this enhanced-SOC regime, the first pair of minibands becomes more clearly isolated and exhibits a robust nontrivial topology, confirmed by the WCC analysis. The increased SOC strengthens the avoided crossings and Berry-curvature concentration inherited from the parent bands, thereby stabilizing the topological minibands and improving their energy isolation.

Guided by the ML-extracted trend and the effective-Hamiltonian verification showing that stronger Dresselhaus SOC improves miniband isolation and topology, we next perform a material search for a chemically realistic way to access this high-SOC regime. Since SOC generally increases with atomic number, replacing Hg with the heavier Pb provides a natural route to enhance SOC while preserving a similar lattice structure, motivating us to consider $PbI_2$. The calculated conduction-band structure of monolayer $PbI_2$, shown in **Fig. 5(b)**, exhibits a pronounced Dresselhaus SOC effect. By fitting the low-energy bands, we extract a Dresselhaus parameter to be 1.02 eV·Å, which is significantly larger than that (0.33 eV·Å) of $HgI_2$. Using the effective Hamiltonian (Eq. (3)) with the extracted parameters of DFT simulations of bilayer $PbI_2$, we have computed the moiré miniband structure, as shown in **Fig. 5(c)**. The WCC simulation illustrated in the inset confirms the nontrivial topology of the second lowest-energy band. Notably, the lowest two mini bands are well isolated, without overlap with other higher-energy bands, making it observable for transport measurement.

Another important consequence of enhanced SOC is that topological minibands become easier to form, thereby reducing the critical twist angle for the topological phase. Since a smaller twist angle typically yields narrower minibands, we analyze the bandwidth evolution as a function of twist angle shown in **Fig. 5(d)**, where the associated topological phase transition is also indicated. We find an exceptionally small bandwidth, below 12 meV, across a relatively broad range of twist angles (2° – 4°), comparable with representative TMD moiré systems such as $MoTe_2$.[30]

Such ultra-narrow bands are expected to enhance electron–electron correlations and may promote correlation-driven magnetism. To check this expectation, we performed Hartree–Fock calculations for fillings corresponding up to four doped electrons per moiré supercell and observed correlation-driven magnetic orders.[64] [see SI for details] As shown in **Fig. 5(e)**, within the most interesting range of the twist angle (from 2.5° to 3.5°), we have compared the total energies of ferromagnetic (FM) and Néel antiferromagnetic (AFM) states, and the AFM state remains as the ground state. We have also checked this conclusion for a reasonably filling range (from 0 to 4), the conclusion of the AFM ground state remains robust (See **Fig. S6**).

Then, we focus on an optimized case for the correlation-induced gap, namely $\theta = 3°$with three-electron doping. Correlation effects dramatically enhance the insulating gap from 0.3 meV to 20 meV, as shown in **Fig. 5(f)**. Such a sizable correlated gap is comparable to those observed in twisted bilayer graphene and twisted TMD systems,[65,66] which greatly improves the feasibility of experimental detection and verification. We also evaluate the topological indicators and find a transition from a trivial to a nontrivial phase in the range 2.8° - 3.4° (see the WCC evolution in **Fig. S4**), indicating an AFM QSH state. Moreover, if the magnetic order is driven from AFM to FM, for example by an external magnetic field, we obtain a Chern-insulating phase with Chern number $C = 1$(**Fig. S5**). These results demonstrate that carrier doping not only induces magnetism but also preserves and enriches the underlying band topology. Consequently, the system provides a tunable route from a QSH phase to a magnetic Chern insulator, offering both a mechanism and a materials platform for realizing and expanding topological phases in twisted systems.

It is worth mentioning again that the above mechanism for realizing moiré topologies is not restricted to tetragonal D-2 systems. It is a general design paradigm for engineering topological minibands and realizing diverse topological phenomena in large-period moiré systems: by stacking two noncentrosymmetric monolayers with opposite polar or symmetry-related orientations to form a type-II Rashba- or Dresselhaus-SOC bilayer, the moiré potential effectively slices and redistributes the Berry curvature, thereby generating topological moiré minibands with correlated effects. Combined with physics-informed ML, our results offer fresh perspectives and practical routes for exploring twistronics and twist-induced strong correlations in general noncollinear SOC materials.

## METHOD

The first-principles simulations are performed using the Vienna ab initio Simulation Package (VASP).[67] The atomic structure optimization is performed within the generalized gradient approximation (GGA) and Perdew–Burke–Ernzerhof (PBE) exchange-correlation functional [68,69] until the force on each atom is less than 0.01 eV/Å. The electronic iteration convergence criterion is set to $1\times10^{-6}$ eV. The k-mesh is set to 15 × 15 × 1 with a 500-eV cutoff energy. A vacuum layer of at least 15 Å was used to avoid interactions between periodic images. Spin–orbit coupling was included in all electronic-structure calculations. The DFT-D3 method is employed to include vdW interactions.[70] The moiré continuum parameters were fitted from high-symmetry stacking configurations of untwisted bilayers.

## AUTHOR INFORMATION

Corresponding Author:

**Li Yang** - Department of Physics and Institute of Materials Science and Engineering, Washington University in St. Louis, St. Louis, Missouri 63130, USA

Email: lyang@physics.wustl.edu

**Shaffique Adam** - Department of Physics and Institute of Materials Science and Engineering, Washington University in St. Louis, St. Louis, Missouri 63130, USA

Email: shaffique@wustl.edu

Author:

**Xilong Xu** - Department of Physics, Washington University in St. Louis, St. Louis, Missouri 63130, USA

Email: xilong@wustl.edu

**Liangtao Peng** - Department of Physics, Washington University in St. Louis, St. Louis, Missouri 63130, USA

Email: liangtao@wustl.edu

## Notes

The authors declare no competing financial interest.

## ASSOCIATED CONTENT

Supporting Information

The Supporting Information is available free of charge at URL.

We provide the results of twisted Type-II Rashba homobilayer BiTeBr, the calculated parameters of bilayer $HgI_2$ and the details of physics-informed ML model and Hartree-Fock theory calculations.

## ACKNOWLEDGMENTS

X.X. is supported by the Department of Energy (DOE), Office of Science, Basic Energy Sciences under Award No. DE-SC0026312. L.Y. is supported by National Science Foundation DMR-2124934. L. P. and S. A. are supported by a start-up grant at Washington University in St. Louis. The simulation used Anvil at Purdue University through allocation DMR100005 from the Advanced Cyberinfrastructure Coordination Ecosystem: Services & Support (ACCESS) program, which is supported by National Science Foundation grants #2138259, #2138286, #2138307, #2137603, and #2138296.

**Figures:**

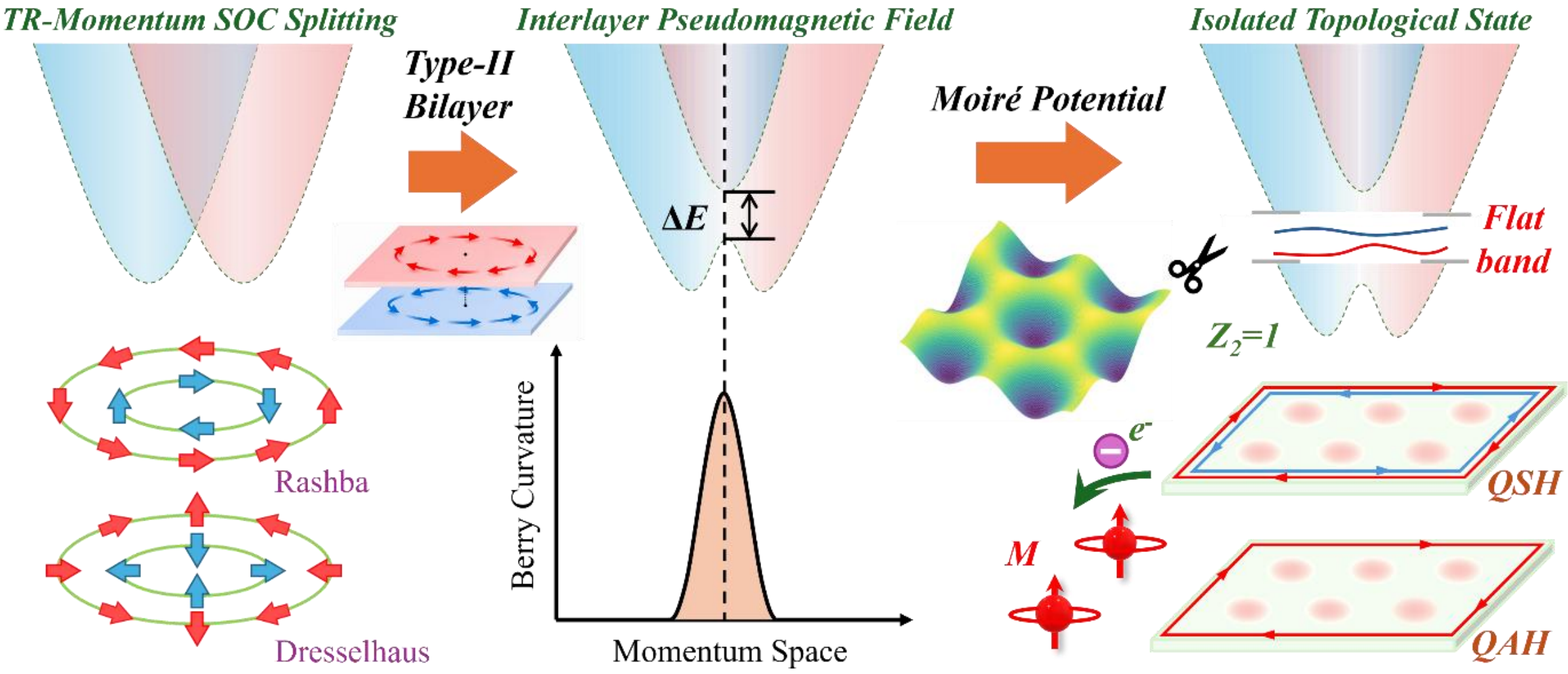

**Figure 1**. Schematic illustration of moiré engineering artificial type-II SOC bilayers to realize topological flat bands and correlated states.

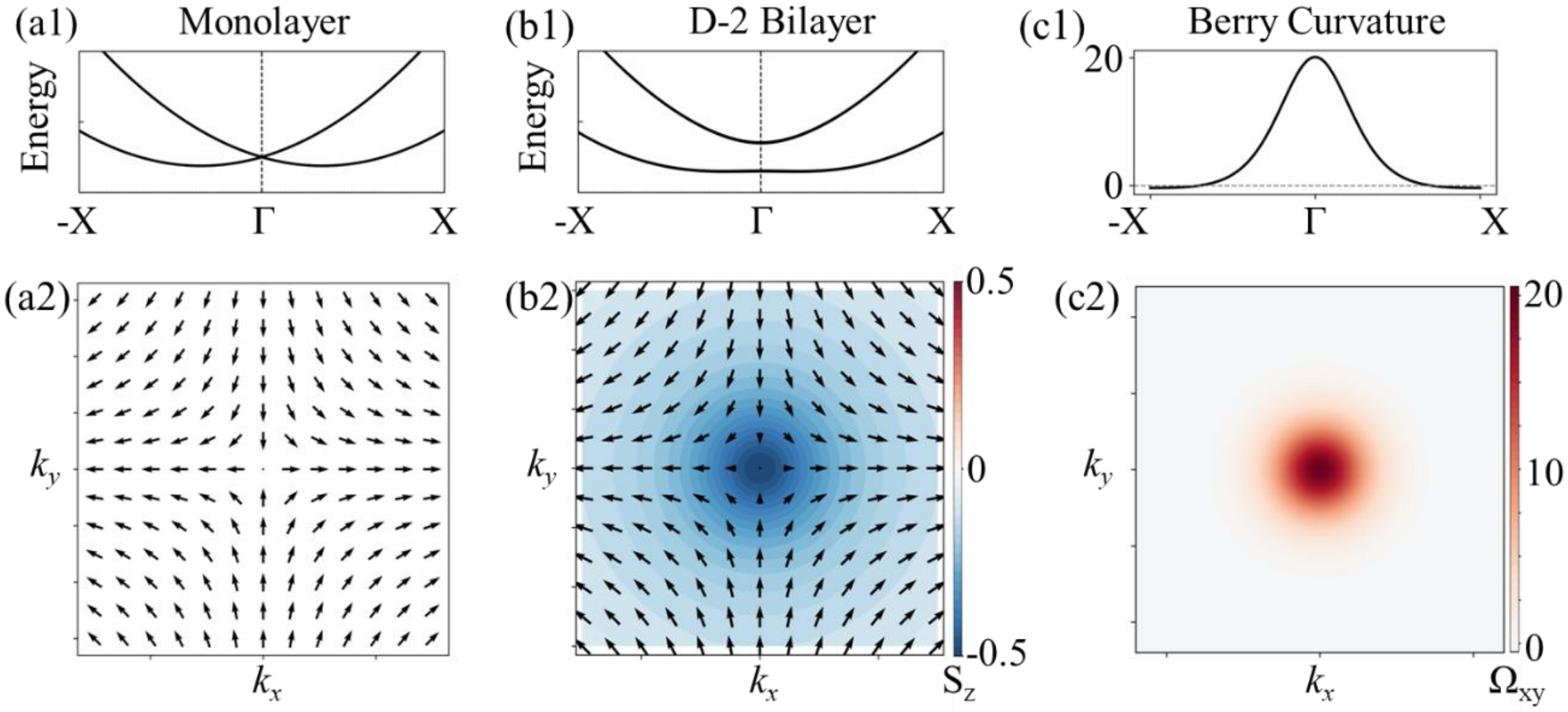

**Figure 2**. Model results of monolayer and D-2 bilayer Dresselhauses-SOC structures. Band structures and spin textures of the effective model in monolayer (a) and bilayer (b) Dresselhauses SOC. (c) Berry curvature around Γ point in D-2 system.

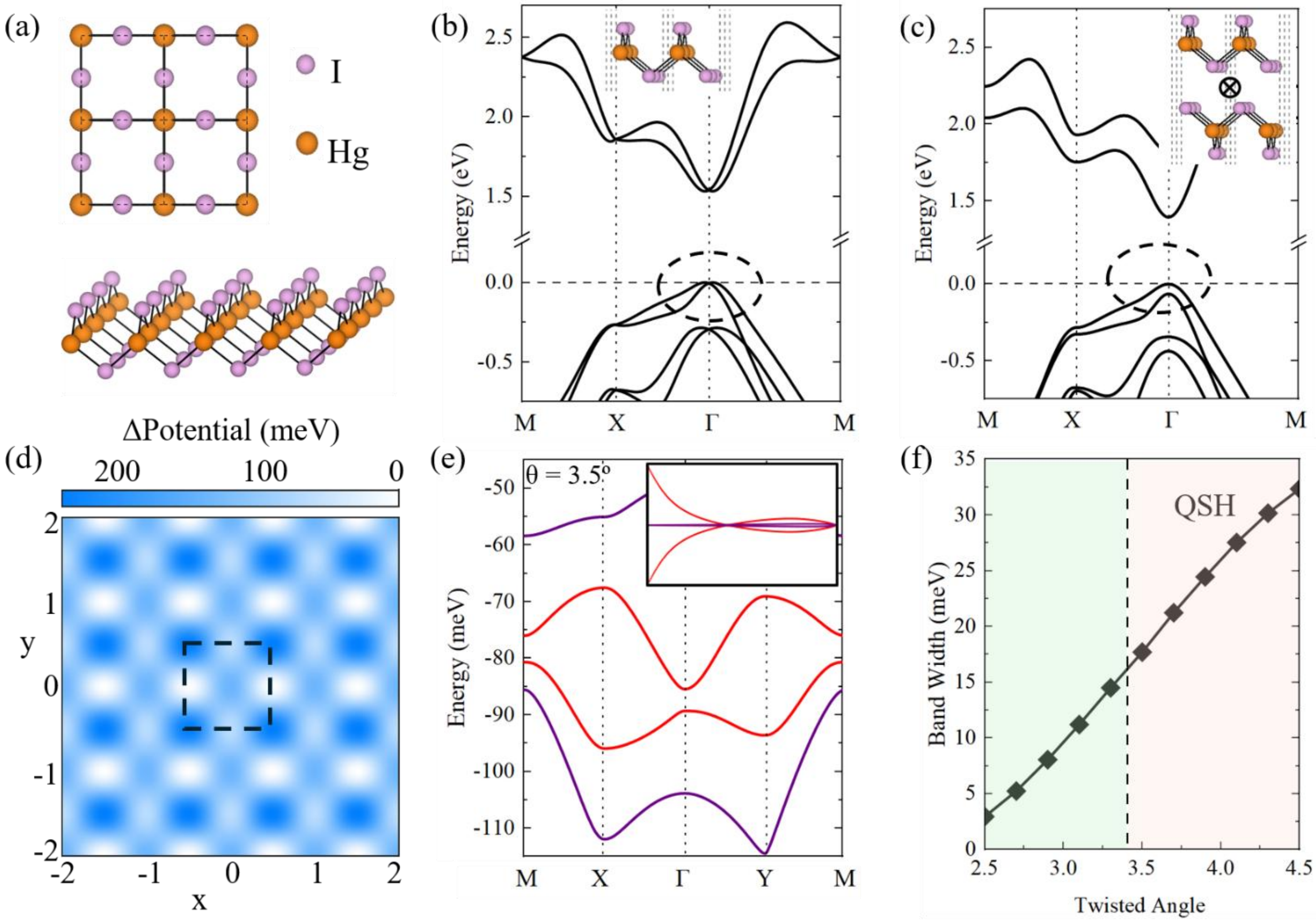

**Figure 3**. Twisted Square-lattice $HgI_2$ moiré structure. (a) Top and side view of monolayer square-lattice $HgI_2$. (b) Electronic band structure of monolayer $HgI_2$, focusing on low-energy states near band edges. (c) Electronic band structure of bilayer $HgI_2$. Inset is the side view of D-2 homobilayer, and the cross symbol marks the inversion center. (d) Moiré potential $\Delta(\boldsymbol{r})$ of twisted D-2 bilayer $HgI_2$. The moiré period along x and y directions is given by $L_M = a/[2\sin(\theta/2)]$, where $a$ is the optimized in-plane lattice constant of monolayer $HgI_2$. This corresponds to a moiré length scale of approximately 7 nm. (e) Moiré miniband dispersion along high-symmetry lines of twisted D-2 bilayer $HgI_2$. Inset is the WCC evolution of the two pairs of the highest-energy valence minibands. (f) Evolution of band width and topological phase transition of the second pair valence moiré bands in twisted D-2 bilayer $HgI_2$.

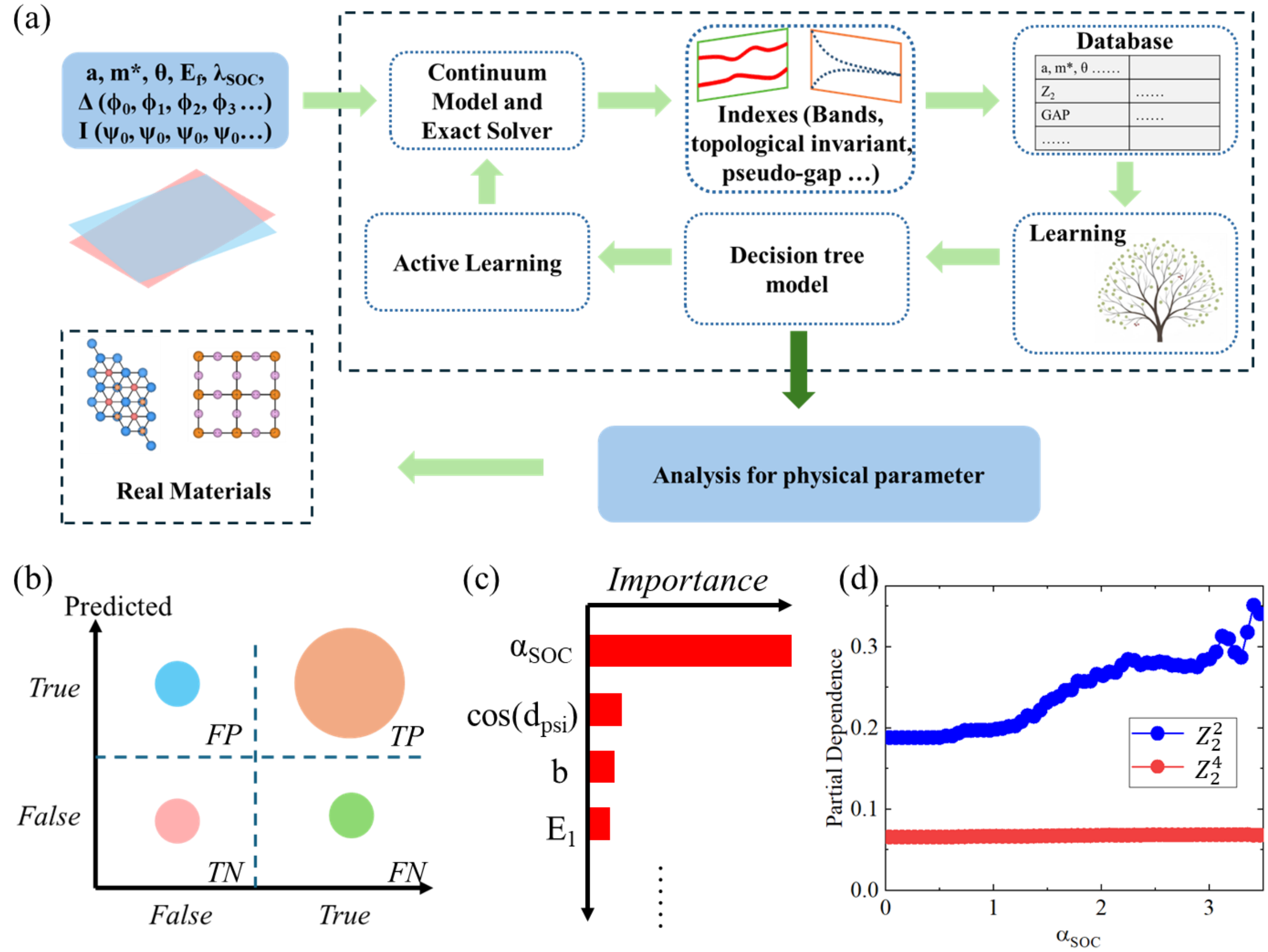

**Figure 4**. Physics-informed ML workflow for twisted SOC moiré systems. (a) Schematic of the physics-informed ML surrogate-model workflow. (b) Illustration of classification performance using a confusion-matrix-style diagram (used together with MCC as the main metric. TP: True Positive, TN: True Negative, FP: False Positive, FN: False Negative). (c) Feature-importance ranking extracted from the trained model, highlighting the dominant role of SOC-related scales. (d) Partial-dependence plot showing how the predicted probability of nontrivial topology (e.g., $Z_2^2$ and $Z_2^4$ at different fillings) varies with $\alpha_{\mathrm{SOC}}$, revealing SOC as the decisive control knob for stabilizing topological minibands.

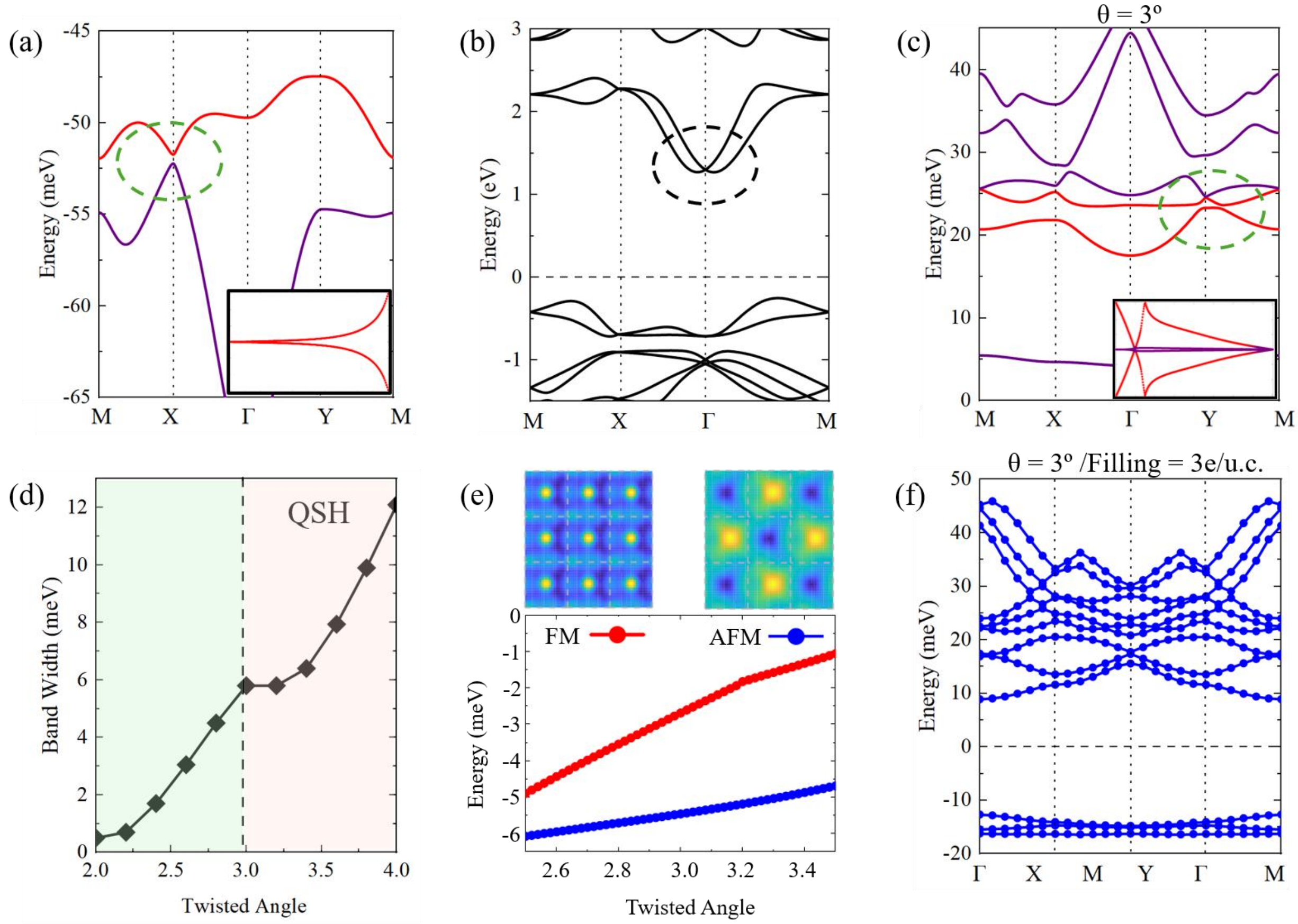

**Figure 5**. Isolated topological bands and correlated effects. (a) Moiré miniband dispersion with artificially larger SOC strength along high-symmetry lines of twisted bilayer $HgI_2$, where band hybridization and avoided crossings emerge (highlighted by red circles). Inset is the WCC evolution. (b) Electronic band structure of monolayer $PbI_2$. (c) Moiré miniband structure of twisted D-2 bilayer $PbI_2$. Inset is the WCC evolution. (d) Evolution of band width and topological phase transition in twisted D-2 bilayer $PbI_2$. (e) Correlated effects in twisted D-2 bilayer $PbI_2$ and spin charge density of FM and AFM twisted D-2 bilayer $PbI_2$. (f) Hartree-Fock band structure of twisted bilayer $PbI_2$ with three-electrons doping.

**Table 1**. Evaluation of the ML model. Acc: Represents the proportion of correctly predicted samples (both positive and negative) among all samples. The formula is defined as $(TP+TN)/(TP+TN+FP+FN)$; BalAcc: Average of sensitivity (true positive rate) and specificity (true negative rate), providing a fairer measure under class imbalance. The formula is defined as 1/2 ("TPR" +"TNR" ), where TPR=$TP/(TP+FN)$, TNR=$TN/(TN+FP)$.

| | **Acc** | **BalAcc** |
|---|---|---|
| $Z_2^2$ | 0.9400 | 0.9419 |
| $Z_2^4$ | 0.7540 | 0.8171 |
| $E_g^{23}$ | 0.9790 | 0.9534 |
| $E_g^{45}$ | 0.9600 | 0.9423 |

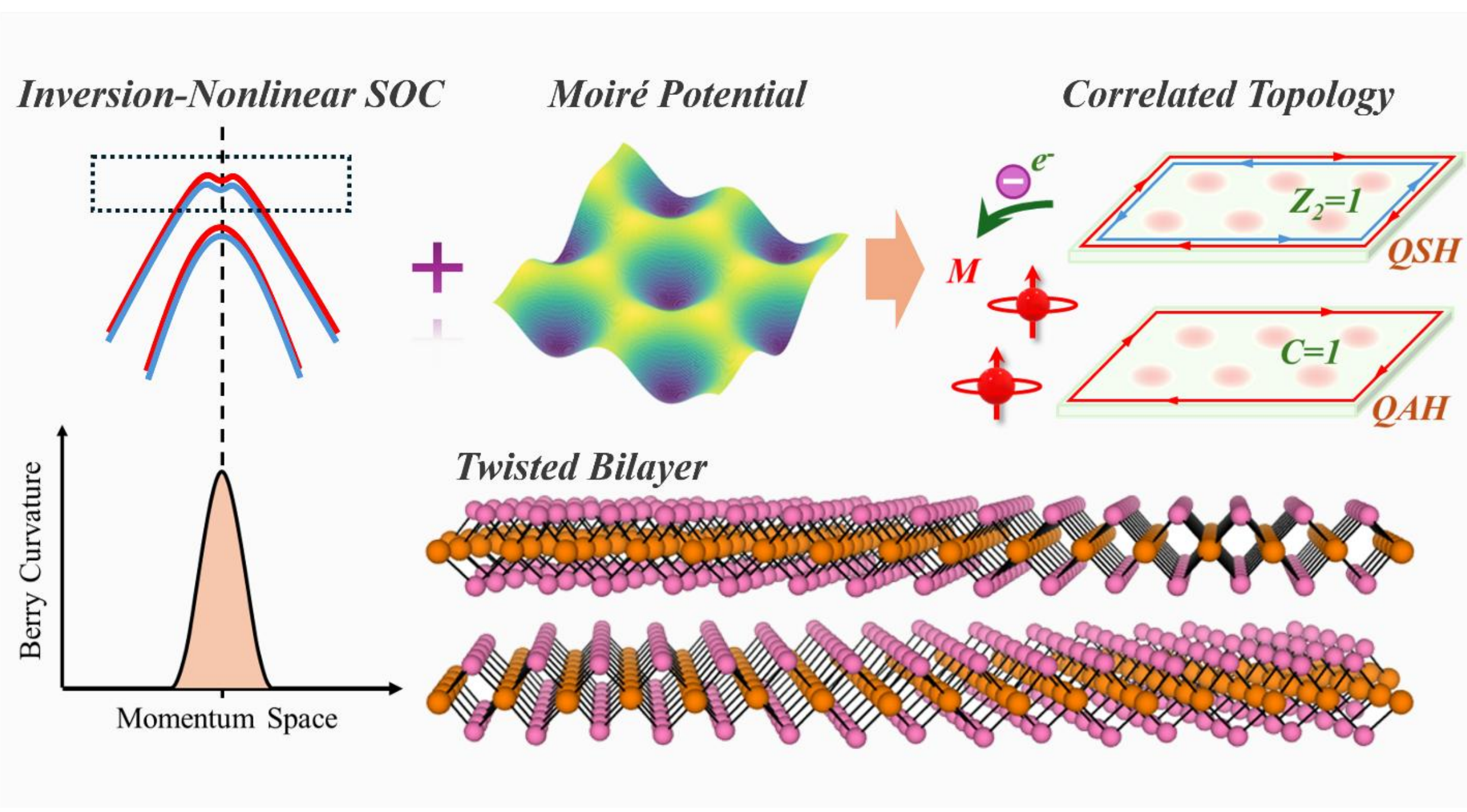
Inversion-Nonlinear SOC
Moiré Potential
Correlated Topology
e-
M
Z2=1
QSH
C=1
QAH
Berry Curvature
Momentum Space
Twisted Bilayer

**TOC Graphic**